\documentclass[a4paper,11pt]{article}
\usepackage{pos}
\usepackage{hyperref}
\usepackage{url}
\usepackage{siunitx}
\usepackage{lineno}
\usepackage{xspace}
\usepackage{bm}
\def\Pp      {\ensuremath{p}\xspace}
\def\proton      {{\ensuremath{\Pp}}\xspace}

\title{Real-time monitoring of LHCb beam spot properties based on FPGA hit reconstruction}

\author*[a,b,d]{Giulio Cordova}
\author[b,d]{Daniele Passaro}
\author[c,d]{Federico Lazzari}
\author[c,d,e]{Elena Graverini}
\author[b,d]{Michael J. Morello}
\author[c,d]{Giovanni Punzi}

\affiliation[a]{CERN, Esplanade des Particules, Geneva, Switzerland}
   
\affiliation[b]{Scuola Normale Superiore, Piazza dei Cavalieri, Pisa, Italy}

\affiliation[c]{Dipartimento di Fisica ``E. Fermi'', Università di Pisa, Largo Bruno Pontecorvo, Pisa, Italy}

\affiliation[d]{INFN Pisa, Largo Bruno Pontecorvo, Pisa, Italy}

\affiliation[e]{École Polytechnique Fédérale, Route Cantonale, Lausanne, Switzerland}

\emailAdd{giulio.cordova@cern.ch}

\abstract{We present a real-time trackless method to monitor the centre
of the LHCb luminous region and the relative alignment of the two
halves of the LHCb Vertex Locator, using cluster counts accumulated in selected regions of the detector and processed in firmware. A Principal Component Analysis performed on simulated data yields a set of weight vectors that reduce the position
estimators to scalar products of the counter rates, executable with
negligible computational overhead. With \proton\proton collision data from 2024, the
transverse position of the luminous region is reconstructed with a
statistical precision of approximately \SI{4}{\micro\meter} per
coordinate in a \SI{90}{\milli\second} accumulation window. A measurement of the longitudinal centre is demonstrated with a statistical precision of a few millimetres in simulated studies. The introduction of run-time configurable firmware
counters has been studied in simulation to improve the transverse and longitudinal
resolutions by a factor of three with respect to the fixed-geometry
baseline.}

\FullConference{23rd International Workshop on Advanced Computing and Analysis Techniques in Physics Research (ACAT2025)\\
8–12 September 2025\\
Hamburg, Germany\\}

\begin{document}
\maketitle

\section{Introduction}
\noindent LHCb is a single-arm forward spectrometer at the LHC, optimised for precision measurements of CP-symmetry violation observables and for the study of rare decays of heavy-flavoured hadrons~\cite{LHCb:2008vvz}. 

For the LHC Run~3, the detector underwent a major upgrade to operate its data processing flow at a rate of \SI{30}{\mega\hertz}~\cite{LHCb:2023hlw}. As part of this upgrade, the silicon-strip VErtex LOcator (VELO) was replaced by a silicon-pixel detector~\cite{Bediaga:2013tje}, arranged in 26 stations along the beam direction; each station consists of two independently movable halves that can be retracted up to \SI{25}{\milli\meter} from the beam axis during injection, and brought as close as \SI{5.1}{\milli\meter} from the beam for data taking.
 
With the upgrade, LHCb adopted a fully triggerless readout in which every colliding bunch crossing is processed: sub-detector data are time-synchronized and assembled to run full event reconstruction in real time.
This involves reconstructing high level objects such as tracks and primary vertices by means of heterogeneous computing systems using FPGAs, GPUs and CPUs~\cite{LHCb:2018mlt,LHCb:2018pqv}. 

As part of the data processing upgrade, a two-dimensional clustering algorithm was designed, that runs directly on the Arria~10 FPGAs of the VELO readout boards~\cite{Bassi_2023,arria10ds}, delivering hit positions at the earliest possible stage of the data-acquisition chain, at an aggregate rate of about $10^{11}$~hits/s. Such an early and abundant stream of hit information opens up possibilities beyond the standard event reconstruction; for example, luminosity determination, detector
diagnostics, alignment, and monitoring of the luminous-region parameters. A new method to measure luminosity in real time has been developed and successfully used throughout the run~\cite{LHCb:2026spn}. Here, we show that using just firmware-reconstructed clusters, without relying on tracks, one can measure the position of the interaction region and the relative position of the VELO halves in real time. 

\section{Real-time cluster counters on the VELO}
\noindent The clustering firmware operates independently on each VELO half-module and outputs the $x,y$ coordinates of the centroid of every reconstructed cluster, which are then incorporated in the raw data stream, and used in the servers performing the reconstruction at the HLT1 level.
 
\begin{figure}
\centering
\includegraphics[width=.3875\textwidth,clip]{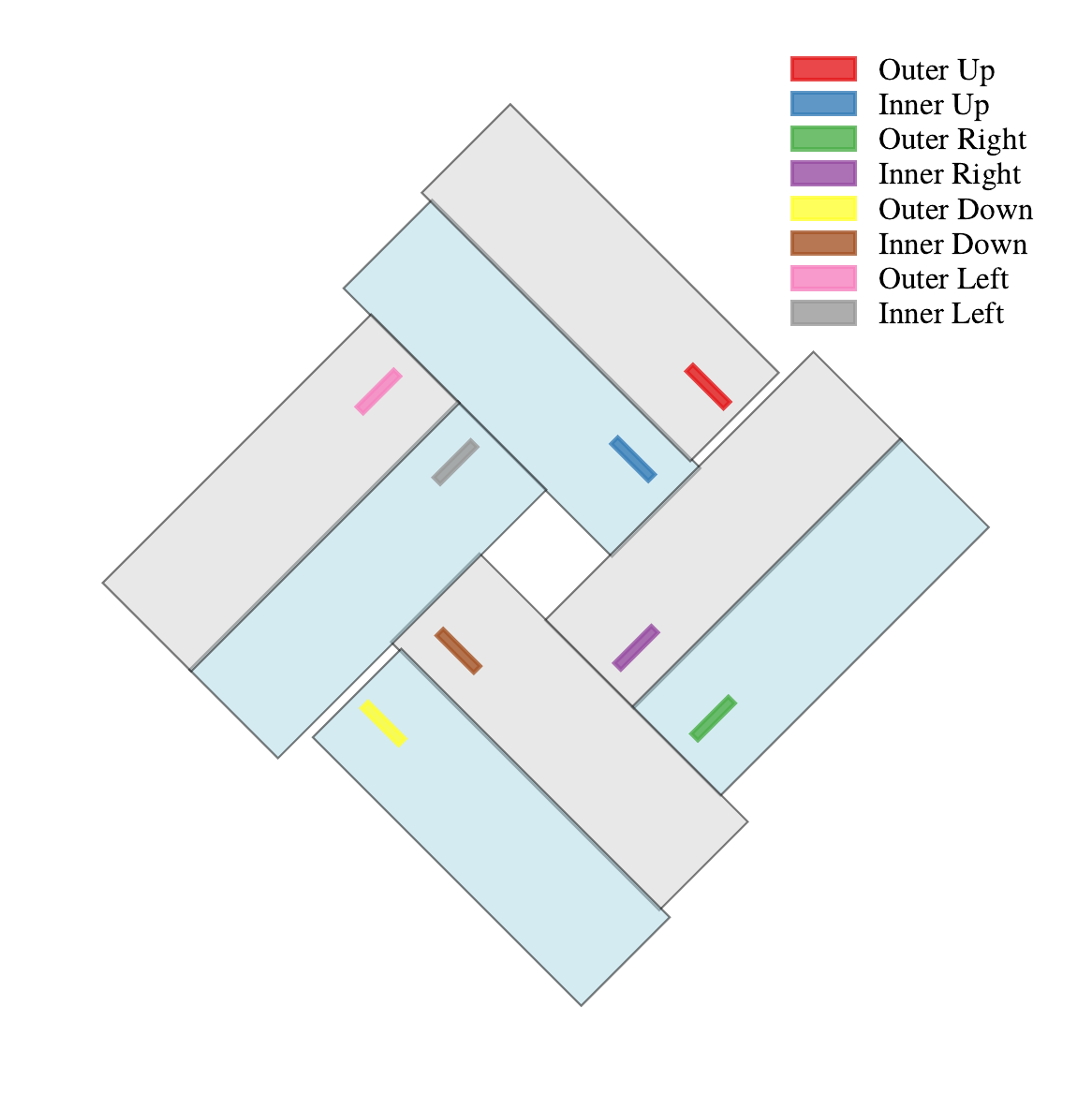}
\includegraphics[width=.3875\textwidth,clip]{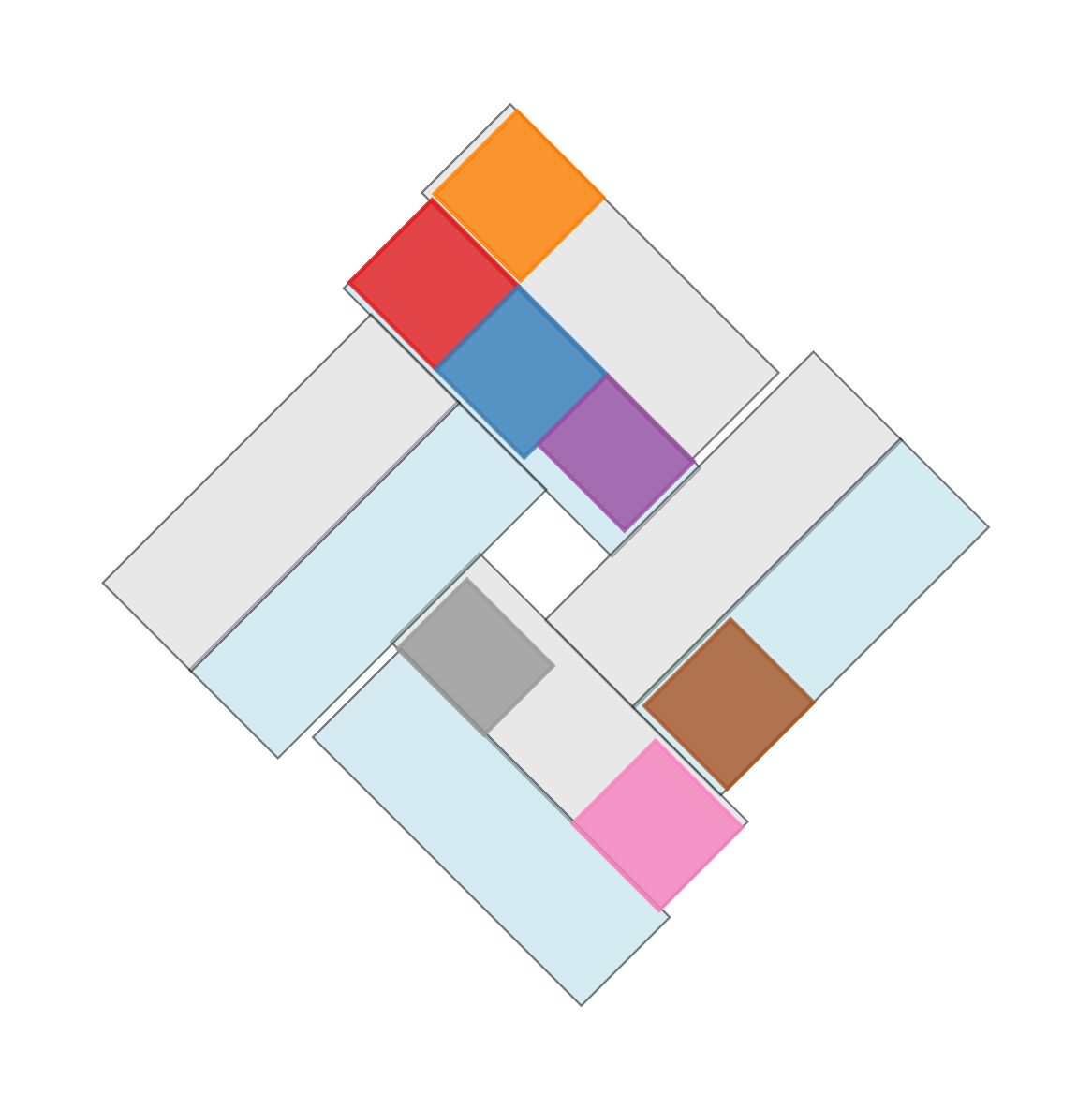}
\caption{(Left) Layout of the fixed cluster-counting regions on one VELO station. Inner and outer regions are shown; within each, four azimuthal sectors (Up, Down, Left, Right) are defined, each covering $20\times110$ pixels on a separate sensor. (Right) Example of programmable counting regions on a single VELO station, illustrating the flexibility in shape and position achievable without firmware recompilation.}
\label{fig:counters}
\end{figure}

Exploiting this stream, two rectangular regions are defined on each VELO sensor at different radial distances from the beam axis, an inner and an outer one; within each of these, four azimuthal sectors -- Up, Down, Left, and Right -- are further identified, each comprising $20\times110$ pixels and located on a different sensor of the station. The geometry, naming convention, and a sketch of these regions on a VELO station are depicted in Figure~\ref{fig:counters}~(left). For each bunch-crossing type (bb, be, eb, ee), clusters recorded in these regions are counted directly in firmware. They are read out and background-subtracted in real time in the LHCb slow-control software, yielding a total of $2\times4\times26=208$ measurements of cluster rates per event. The number of \proton\proton collision events integrated to calculate this rate is a flexible parameter that can be changed in the LHCb slow-control software; it is set by default to the number of beam crossings in 1024 LHC beam orbit revolutions.

The resource footprint of this counting logic is modest: about 0.6\% of the FPGA logic elements, 2.4\% of the M20K memory blocks, and 2.8\% of the digital signal processors. 
 
In addition to these fixed counters, a set of programmable counters has been implemented, configurable at runtime without the need to recompile the FPGA firmware. Up to three such counters can be defined per VELO sensor, each covering an arbitrary rectangular region. An example of 7 different flexible regions in a single module is depicted in Figure~\ref{fig:counters}~(right).
This flexibility allows the integrated area -- and hence the acceptance -- to be tuned for specific applications. For example, an accumulation region can be set up to integrate hits reconstructed even on a whole VELO ASIC, maximizing the rate of clusters and improving the statistical precision of the resulting estimators.

\section{From counter rates to luminous-region parameters}
\label{sec:method}
\noindent 
The information collected by the cluster counters is combined within the LHCb slow-control software to build real-time estimators of both the luminosity and the position of the interaction region. 
The latter are obtained from the counter stream by means of scalar products of the one-padded $(n+1)$-dimensional vector $\mathbf{c}$ of all the available $n$ counters with vectors of appropriate coefficients $\mathbf{w}_{x,y,z}$:
\begin{equation}
\hat{x}=\hat{x}\!\left(1,c_1,\dots,c_n\right)=\mathbf{w_x}\cdot \mathbf{c}, \qquad
\hat{y}=\hat{y}\!\left(1,c_1,\dots,c_n\right)=\mathbf{w_y}\cdot\mathbf{c}, \qquad
\hat{z}=\hat{z}\!\left(1,c_1,\dots,c_n\right)=\mathbf{w_z}\cdot\mathbf{c}.
\label{eq:xhat1}
\end{equation}
The one-padding is used to absorb a potential intercept term directly in the scalar product.
The coefficient vectors $\mathbf{w}_{x,y,z}$ used in this work are obtained from Monte Carlo (MC) simulated samples by means of a Principal Component Analysis (PCA)~\cite{Pearson01111901}.
The PCA is a standard technique for dimensionality reduction and feature extraction, based on an orthogonal linear transformation that re-expresses a dataset along directions of decreasing variance. 
In our case, the observables are a set of $n$ counters that can be mapped in any $l$-dimensional space, with $l\leq n$.

This set of Principal Components is then fitted to the desired variable, resulting in the closed formulas
\begin{equation}
\hat{x}=\mathbf{\alpha_x}\cdot \mathbf{t_x} =\mathbf{\alpha_x} A_x \mathbf{c} = \mathbf{w_x} \cdot\mathbf{c} , \qquad
\hat{y}=\mathbf{\alpha_y}\cdot \mathbf{t_y} =\mathbf{\alpha_y} A_y \mathbf{c}= \mathbf{w_y} \cdot\mathbf{c}, \qquad
\hat{z}=\mathbf{\alpha_z}\cdot \mathbf{t_z} =\mathbf{\alpha_z} A_z \mathbf{c}= \mathbf{w_z} \cdot\mathbf{c},
\label{eq:xhat2}
\end{equation}
where $\mathbf{\alpha_{x,y,z}}$ are the fit coefficients vectors, and $A_{x,y,z}$ the PCA matrices transforming the original space in the latent one using e.g. $\mathbf{t_x} = A_x\mathbf{c}$.  
The weight vectors $\mathbf{w_{x,y,z}} =A_{x,y,z}^{\mathsf{T}}\mathbf{\alpha_{x,y,z}}$  are obtained from dedicated training samples $\mathbf{X}_{\rm train}$, $\mathbf{Y}_{\rm train}$, and $\mathbf{Z}_{\rm train}$ generated with the LHCb simulation by displacing the simulated beam spot along $x$, $y$, $z$, respectively. First, a PCA is performed on each data sample individually to get the matrices $A_{x,y,z}$ and the components $\mathbf{t_{x,y,z}}$ are fitted to get the coefficients $\mathbf{\alpha_{x,y,z}}$. The final form of the estimators is the desired scalar product, which is a simple quantity that can be computed quickly without relying on extensive resources. 

Since the luminous-region position is, by construction, the only varying feature of these samples, more than 90\% of their variance is explained by the leading component alone, with subleading components dominated by statistical fluctuations; the leading score is therefore expected to be approximately proportional to the corresponding luminous-region coordinate and is the only one retained for $x$ and $y$. This expectation is confirmed on an independent test sample, on which the leading score is found to depend linearly on the simulated shift of the luminous region over the relevant range~\cite{proceeding-chep}.

\begin{figure}
\centering
\includegraphics[width=.4875\textwidth,clip]{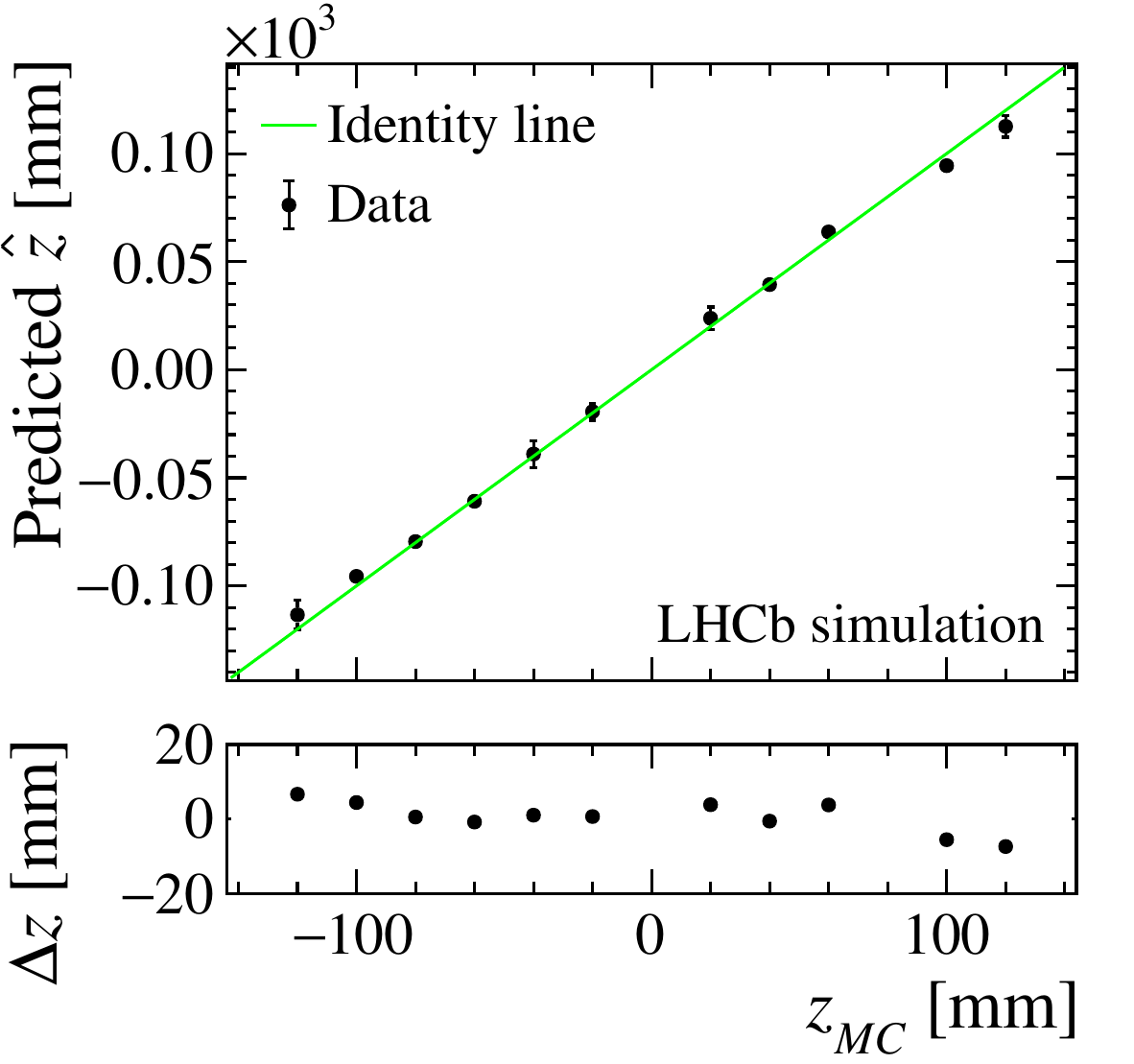}\hfil
\includegraphics[width=.4875\textwidth,clip]{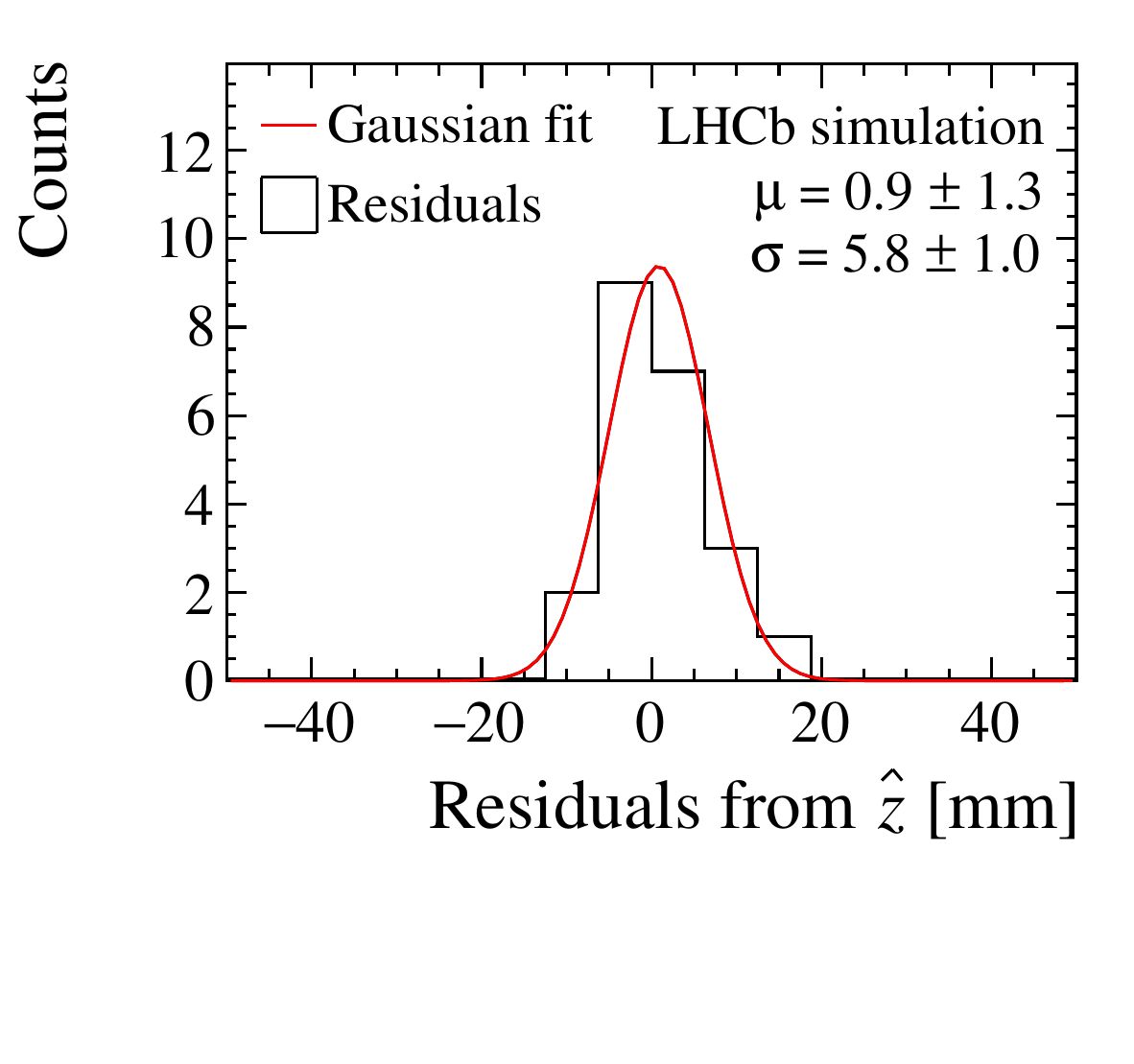}\hfil
\caption{Performance of the longitudinal position estimator on an independent MC test sample. (Left) Estimated $z$ position versus the true simulated value; the green line indicates the identity. The estimator is approximately linear over the range expected during normal operations, with a residual cubic trend at extreme displacements well outside the operational beam position.
(Right) Distribution of residuals, fitted with a Gaussian; the result confirms that the estimator is unbiased and achieves a resolution of a few millimetres
throughout the accessible range.}
\label{fig:z_estimator}
\end{figure}

 The same approximation does not hold along $z$, where the range spanned by the luminous region is comparable to the detector size. In this case, the leading PC alone explains only about 70\% of the variance of the training sample $\mathbf{Z}_{\rm train}$, significantly less than for the transverse coordinates, and its score $t_1^z$ exhibits an approximately cubic dependence on the true longitudinal position rather than a linear one. Using a single component would therefore result in a strongly biased estimator over the full accessible range. Including 30 subleading components, which carry the residual variance associated with higher-order position dependence, and combining them linearly as in Eq.~\eqref{eq:xhat2} produces an estimator that with good linearity over the range expected during normal operations. The performance on an independent test sample is shown in Figure~\ref{fig:z_estimator}: the estimated position follows the true MC value closely in the central region, with a residual cubic trend only at the extreme displacements, well outside the beam positions expected during physics fills. The residuals follow a Gaussian distribution throughout, with no significant bias and a resolution of a few millimetres, which is adequate for the intended monitoring application.

With the same amount of statistics, it is also possible in MC to compare the results using only the inner and outer (small) counters with the flexible ones that cover all VELO sensors. For the transverse components, the resolution of \SI{20}{\micro\meter} obtained with the small counters improves to  \SI{6}{\micro\meter}, while the $z$ component resolution improves from  \SI{6}{\milli\meter} to  \SI{2}{\milli\meter}.

\section{Results on 2024 proton-proton collision data}
\noindent
Since the computation of the scalar product requires all the counters to be available, the procedure has been strengthened against missing or anomalous channels: at each time $\tau$, the median value $m^r(\tau)$ of the counters of a given category ($r=$~inner or outer) is computed, and any counter that departs from $m^r(\tau)$ by more than 50\% is treated as an outlier and replaced by the median before evaluating the score. This keeps the estimator well-defined even when individual channels are temporarily unavailable or affected by detector or readout glitches. 
 
The VELO Closing Monitoring Tasks (VCMT), which reconstruct primary vertices from VELO tracks with a few-millisecond cadence, are used as a reference to compare the position estimates. The length-scale calibration (LSC) phase of a short van der Meer scan~\cite{Balagura:2020fuo}, performed on 6~April~2024, is used for this purpose: during an LSC, the two beams are displaced head-to-head by known steps, producing a shift of the luminous-region position large enough to calibrate and test the estimators.
 
Using a 90~ms accumulation window, the relative resolution with respect to the VCMT is about 4~$\mu$m for both transverse coordinates. Under nominal Run~3 $pp$ running conditions, with higher pile-up and more populated bunches, the same statistical precision is reached with data collected in about 1~ms~\cite{proceeding-chep}.
The same procedure, applied using only the counters of a single VELO half, provides an estimate of the position of that half relative to the luminous region, or equivalently of the relative alignment between the two halves. In this configuration, a resolution of about 4~$\mu$m is obtained along $y$ and about 7~$\mu$m along $x$ for both VELO sides~\cite{proceeding-chep}.

\section{Conclusions}
\noindent We have presented a real-time, trackless method to monitor the position of the LHCb luminous region and the relative alignment of the VELO halves, based solely on cluster counts accumulated in the readout boards of the VELO detector. Using a set of counters and a PCA-derived set of weights pre-computed on simulated data, the estimators reduce at runtime to a single scalar product per coordinate. 
With 2024 $pp$ collision data, the method achieves a statistical resolution of approximately \SI{4}{\micro\meter} on each transverse coordinate in a \SI{90}{\milli\second} accumulation window; under nominal Run~3 bunch-crossing and pile-up conditions, the same precision is expected in about \SI{1}{\milli\second}. Restricting the analysis to a single VELO half yields resolutions of \SI{4}{\micro\meter} along $y$ and \SI{7}{\micro\meter} along $x$, providing a direct handle on the relative position of the two detector halves with respect to the luminous region.
A longitudinal estimator, based on a linear combination of multiple principal components to compensate for the non-linear response of the leading score along $z$, achieves a resolution of a few millimetres in MC studies over the full simulated position range of the luminous region. Monte Carlo studies further show that replacing the fixed $20\times110$-pixel counting regions with the newly implemented programmable counters, covering a larger fraction of each sensor, improves the transverse resolution from \SI{20}{\micro\meter} to \SI{6}{\micro\meter} and the longitudinal resolution from \SI{6}{\milli\meter} to \SI{2}{\milli\meter} for the same integrated statistics.
These results demonstrate that FPGA-level hit reconstruction, combined with simple real-time arithmetic in the slow-control layer, can provide reconstruction-quality spatial information at a very early stage of the data-acquisition chain, with potential applications to luminosity determination, detector alignment monitoring, and beam diagnostics.
\bibliographystyle{JHEP}
\bibliography{main}

\end{document}